\documentclass[amsmath,twocolumn,amssymb,aps,superscriptaddress,showpacs,floatfix,letterpaper,prl,10pt]{revtex4-1}

\usepackage{graphicx}
\usepackage{dcolumn}
\usepackage{bm}
\usepackage{amsmath}
\usepackage{hyperref}

\renewcommand{\H}{\mathcal{H}}
\newcommand{\ket}[1]{\left| #1 \right\rangle}
\newcommand{\bra}[1]{\left\langle #1 \right|}
\newcommand{\eps}{\varepsilon}
\newcommand{\E}{\mathrm{e}}
\newcommand{\I}{\mathrm{i}}
\renewcommand{\Re}{\mathrm{Re}}

\begin{document}

%\title{How to distinguish specular from retro Andreev reflection in graphene rings} % original (arXiv) title
\title{How to Distinguish between Specular and Retroconfigurations for Andreev Reflection in Graphene Rings} % PRL title

\author{J{\"o}rg Schelter}

\author{Bj{\"o}rn Trauzettel}
\affiliation{Institute for Theoretical Physics and Astrophysics,
University of W{\"u}rzburg, 97074 W{\"u}rzburg, Germany}

\author{Patrik Recher}
\affiliation{Institute for Theoretical Physics and Astrophysics,
University of W{\"u}rzburg, 97074 W{\"u}rzburg, Germany}
\affiliation{Institute for Mathematical Physics, TU Braunschweig, 38106 Braunschweig, Germany}

\date{\today}

\begin{abstract}
%We present numerical transport calculations of Andreev reflection in a graphene ring system which is threaded by a magnetic flux and attached to one normal conducting and one superconducting lead. To this end, we solve the Bogoliubov-de Gennes equation for the tight binding model using the recursive Green's functions technique within the Landauer-B\"uttiker framework for elastic transport. By tuning chemical potential and bias voltage, it is possible to switch between regimes where electron and hole originate from the same band (\emph{retro configuration}) or from different bands (\emph{specular configuration}) of the graphene dispersion, respectively. We find that the dominant contributions to the Aharonov-Bohm oscillations in the subgap transport are of period $h/2e$ in retro configuration, whereas in specular configuration they are of period $h/e$. This result confirms the predictions obtained from a qualitative analysis of interfering scattering paths, and since it is robust against disorder and moderate changes of the system, it provides a clear signature to distinguish both types of Andreev reflection processes in graphene. % original (arXiv) abstract
We numerically investigate Andreev reflection in a graphene ring with one normal conducting and one superconducting lead by solving the Bogoliubov--de Gennes equation within the Landauer-B\"uttiker formalism. By tuning chemical potential and bias voltage, it is possible to switch between regimes where electron and hole originate from the same band (retroconfiguration) or from different bands (specular configuration) of the graphene dispersion, respectively. We find that the dominant contributions to the Aharonov-Bohm conductance oscillations in the subgap transport are of period $h/2e$ in retroconfiguration and of period $h/e$ in specular configuration, confirming the predictions obtained from a qualitative analysis of interfering scattering paths. Because of the robustness against disorder and moderate changes to the system, this provides a clear signature to distinguish both types of Andreev reflection processes in graphene. % PRL abstract
\end{abstract}

\pacs{72.80.Vp, 73.23.-b, 74.45.+c, 85.35.Ds}

\maketitle

Since its first experimental realization in 2004~\cite{Novoselov22102004}, graphene has strongly influenced the field of mesoscopic physics due to the peculiar behavior of its electronic excitations as massless Dirac fermions. Two prominent features, namely the effects of Klein tunneling~\cite{Katsnelson:2006,*PhysRevB.74.041403} and Andreev reflection~\cite{PhysRevLett.97.067007}, concern processes at interfaces between regions of different doping and normal metal-superconductor (NS) junctions, respectively (for a review see Ref.~\cite{RevModPhys.80.1337}). While both effects are known to be closely related~\cite{PhysRevB.77.075409}, different aspects of Klein tunneling have become experimentally accessible in the last years~\cite{PhysRevLett.102.026807,*Young:2009}, whereas specular Andreev reflection has not been observed to date, although there exist a number of proposals for the experimental control~\cite{PhysRevLett.103.167003} and detection~\cite{PhysRevLett.97.067007,PhysRevB.75.045417,*PhysRevB.83.235403} of this process. In this Letter, we will present a novel approach concerning the identification of specular Andreev reflection, distinguishing it from conventional retroreflection, and discuss the advantages over previous works in the field.

\begin{figure}[t]
	\centering
	\includegraphics[width=0.95\columnwidth]{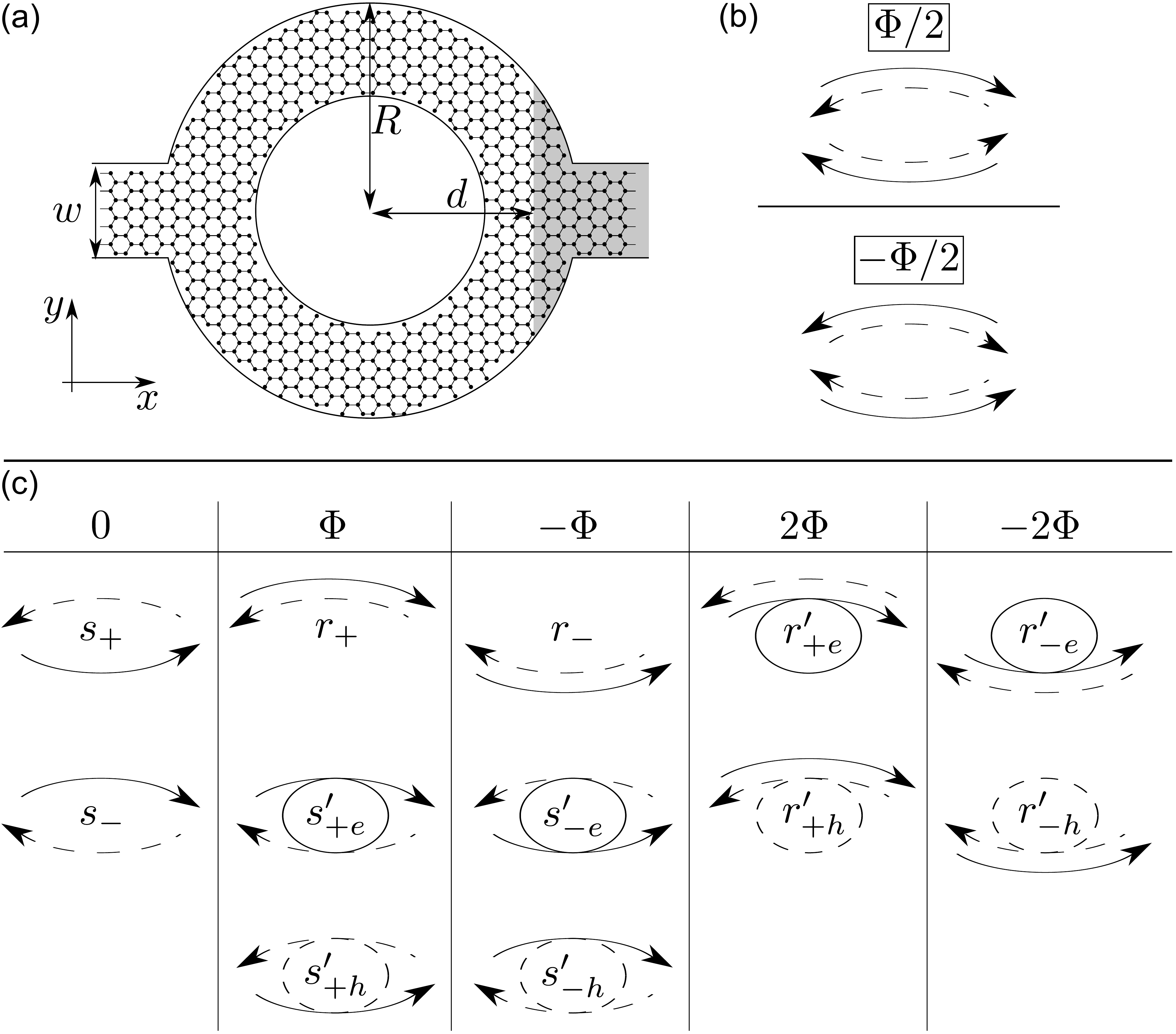}
	\caption{(a) Device geometry showing a graphene ring structure that is penetrated by a magnetic flux $\Phi$ measured in units of the flux quantum $\Phi_0$. At the interface with the superconductor (shaded region), electron-hole conversion may occur. (b) The gauge is chosen such that each of the eight individual electron (solid lines) and hole (dashed lines) paths picks up a phase $\pm \Phi/2$ as indicated. (c) Scattering paths for electrons injected from and holes leaving through the left normal conducting lead; only zeroth and first order contributions are included, i.\,e.\ terms containing a single electron-hole conversion process and at most one additional round-trip of the electron or the hole. The paths are categorized according to the total phase that is picked up, and each path is associated with a corresponding amplitude, where first order amplitudes are indicated by a prime.}
	\label{fig:paths}
\end{figure}

Our approach is based on the observation that in general, the probability for an incident electron to be reflected as a hole is less than one. This allows for effects typical for phase-coherent mesoscopic devices, like universal conductance fluctuations or Aharonov-Bohm oscillations~\cite{PhysRev.115.485} in the magnetoconductance. While in normal metals, the fundamental period of these oscillations is given by the flux quantum $\Phi_0 = h/e$, it is half the value for Andreev (retro)reflection in conventional metals, due to the charge $2e$ of a Cooper pair. However, this is not true anymore in the case of specular Andreev reflection, therefore providing a criterion to distinguish between specular and retroreflection. In order to show this, we consider the phases due to the magnetic flux that are picked up by the various scattering paths. In this analysis, we restrict ourselves to the contributions up to first order in the sense that we take processes into account that involve only a single electron-hole conversion process, and that contain at most one additional round-trip of electron or hole, respectively; higher order contributions connected with additional round-trips are often times negligible~\cite{PhysRevB.81.195441,PhysRevB.77.085413}. The corresponding 
paths are summarized in Fig.~\ref{fig:paths}. In order to obtain the magnetoconductance for the two types of Andreev reflection [specular ($s$) and retro ($r$)], we sum up the amplitudes as defined in Fig.~\ref{fig:paths} for the various paths coherently to obtain the corresponding Andreev reflection probabilities:
\begin{align}
	R_s(\Phi) \cong& \left| 
	s_+ + s_- + s_+' \E^{\I \Phi} + s_-' \E^{-\I \Phi}
	\right|^2 \nonumber \\
	R_r(\Phi) \cong& \left| 
	r_+ \E^{\I \Phi} + r_- \E^{-\I \Phi} + 
	r_+' \E^{2 \I \Phi} + r_-' \E^{-2 \I \Phi}
	\right|^2
\end{align}
where $s_\pm' = s_{\pm e}' + s_{\pm h}'$, $r_\pm' = r_{\pm e}' + r_{\pm h}'$, and $\Phi$ is the magnetic flux measured in units of the flux quantum $\Phi_0$. 
Assuming $|s| \gg |s'|$ for any zeroth- and first-order amplitudes, respectively, we obtain
\begin{equation}
\label{for:Ts}
	R_s(\Phi) \cong R_s^0 + 
	2 \Re \left[(s_+' s_0^* + s_0 s_-'^*)  \E^{\I \Phi}\right] + 
	\mathcal{O}[(s')^2],
\end{equation}
where $s_0 = s_+ + s_-$ and $R_s^0$ contains contributions that are constant with respect to $\Phi$. Therefore, in the case of specular reflection, oscillations of period $h/e$ are dominant. In contrast, in the case of retroreflection, contributions of period $h/2e$ are dominant, as expected:
\begin{equation}
\label{for:Tr}
	R_r(\Phi) \cong R_r^0 +
	2 \Re \left[r_+ r_-^* \E^{2 \I \Phi}	\right] + \mathcal{O}[r r', \ (r')^2],
\end{equation}
where again $R_r^0$ contains $\Phi$-independent terms and we assume $|r| \gg |r'|$ for any zeroth- and first-order amplitudes, respectively.

In order to test this analysis on the basis of a microscopic model, we implement the Bogoliubov-de Gennes Hamiltonian~\cite{gennes1999superconductivity}
\begin{equation}
\label{for:hamBdG}
	H = \left(
	\begin{array}{cc}
		\H	 - E_F			& \boldsymbol{\Delta}\\
		\boldsymbol{\Delta}				& E_F - \H^*
	\end{array}
	\right)
\end{equation}
within the tight binding formalism of graphene
\begin{equation}
\label{for:hamTB}
	\H = \sum_i U_i \ket{i} \bra{i} + 
	\sum_{\left\langle i, j \right\rangle} \tau_{ij} \ket{i} \bra{j}
\end{equation}
where the second sum runs over nearest neighbors and $U_i = U(\mathbf{r}_i)$ is a position-dependent potential. In this numerical calculation, all higher order contributions are also taken into account. 
In Eq.~\eqref{for:hamBdG}, we assume $\Delta_i = \Delta(\mathbf{r}_i) \in \mathbb{R}$ for the superconducting order parameter $\boldsymbol{\Delta} = \sum_i \Delta_i \ket{i} \bra{i}$. The presence of a magnetic field is captured by a Peierl's phase in the hopping matrix element
\begin{equation}
	\tau_{ij} = -\tau_0 \exp{ \left( \frac{ 2 \pi \mathrm{i} }{ \Phi_0 } \int_{ \mathbf{r}_i }^{ \mathbf{r}_j }{ \mathbf{A}(\mathbf{r}) \mathrm{d}\mathbf{r} } \right) },
\end{equation}
where $\tau_0 \approx 2.7$~eV is the graphene hopping integral, $\Phi_0 = h / e$ is the magnetic flux quantum, and the line integral is taken along the straight path between sites $i$ and $j$.

The structure of the graphene device under consideration is schematically shown in Fig.~\ref{fig:paths}. The ring-shaped structure is generated by setting the appropriate hopping matrix elements to zero in Eq.~\eqref{for:hamTB}. The two semi-infinite leads also exhibit the graphene lattice structure; superconductivity is induced into the right lead due to the proximity effect of a superconducting electrode on top of the graphene. We choose to orient the leads to exhibit armchair edges and later comment on the reason for this particular choice. The whole ring is penetrated by a uniform perpendicular magnetic field of strength $B$, described by the vector potential $\mathbf{A}(\mathbf{r}) = -B y \theta(d - |x|) \mathbf{\hat{e}}_x$. The origin of coordinates is taken at the center of the ring.

\begin{figure}[b]
	\centering
	\includegraphics[width=0.95\columnwidth]{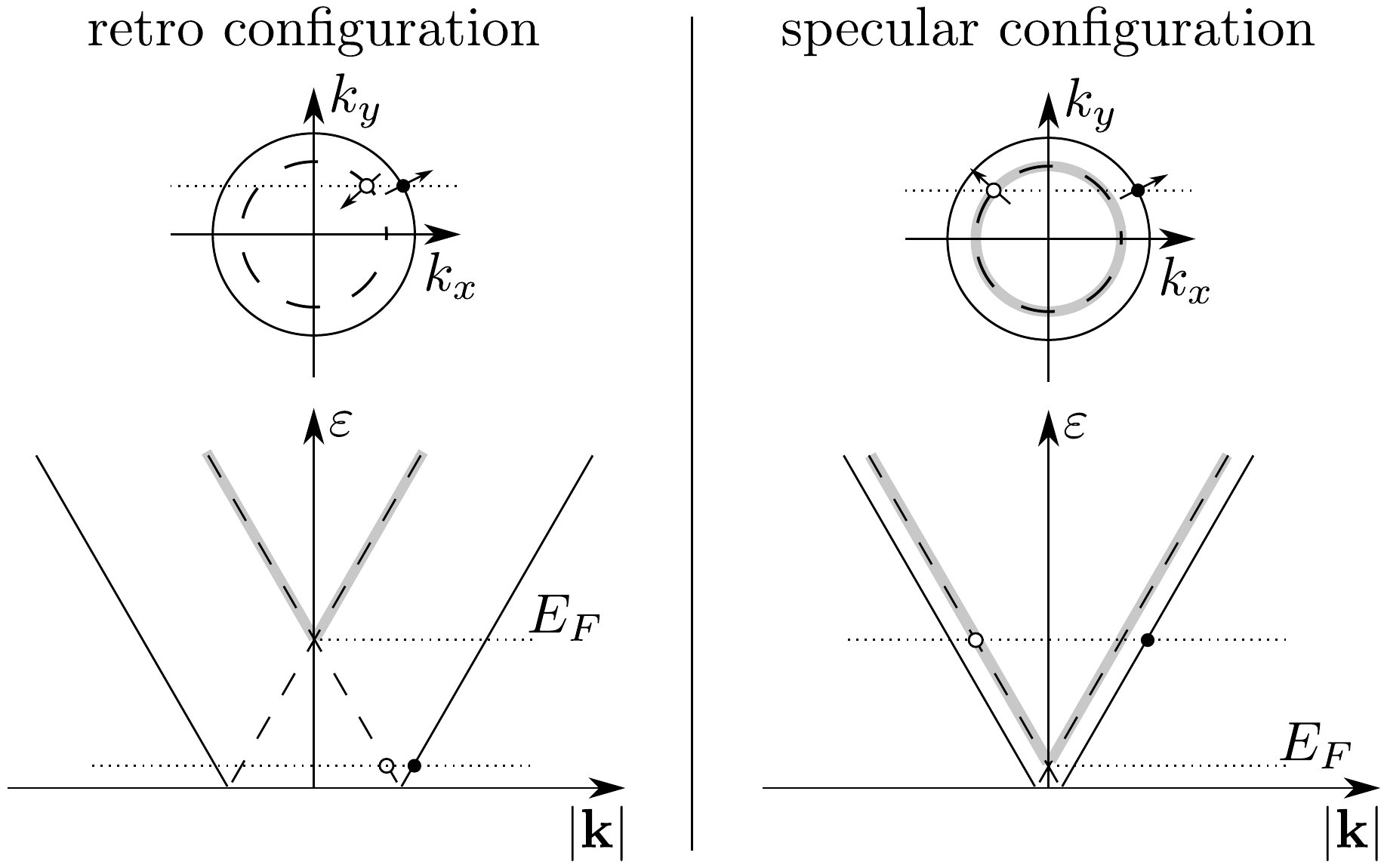}
	\caption{Schematics of the excitation spectrum (lower panel) and surfaces of constant excitation energy in $\mathbf{k}$ space (upper panel) in the cases $E_F > \eps > 0$ (retroconfiguration) and $0 < E_F < \eps$ (specular configuration). Solid and dashed lines indicate electron- and hole-like states, respectively, (hole) states originating from the valence band are shaded gray. The small arrows in the upper panel indicate the direction of propagation of the corresponding states. Electron-hole excitations are drawn assuming conservation of $k_y$ at the NS interface.}
	\label{fig:dispersion}
\end{figure}

In order to fulfill the mean-field requirement of superconductivity, which demands the superconducting coherence length $\xi = \hbar v_F / \Delta$ to be large compared to the wavelength $\lambda_S$ in the superconducting region~\cite{PhysRevLett.97.067007}, we introduce additional doping into the superconducting region by applying a gate potential $U_i = U \theta(x_i - d)$. Which type of Andreev reflection occurs at the NS interface is then determined by the excitation energy $\eps$ [i.\,e.\ the eigenvalues of Eq.~\eqref{for:hamBdG}] and the Fermi energy $E_F$, as shown in Fig.~\ref{fig:dispersion}. In retroconfiguration, $E_F > \eps > 0$, where $v_y^{(h)} \cdot v_y^{(e)} < 0$ for the $y$-components of the electron and hole velocities, both electron and hole traverse the same arm of the ring. In specular configuration, $0 < E_F < \eps$, the hole is reflected back through the other arm of the ring, since $v_y^{(h)} \cdot v_y^{(e)} > 0$. In the following, we choose $|U| \gg E_F$, justifying the adoption of the step-function model for the superconducting order parameter, $\Delta_i = \Delta \theta(x_i - d)$~\cite{PhysRevLett.97.067007}.

In order to compare retro ($r$) and specular ($s$) configurations, we will choose $\eps^{(r)} = E_F^{(s)}$ and $\eps^{(s)} = E_F^{(r)}$ since then the states in both configurations exhibit the same wavelength and there is the same number of propagating modes. We further choose $\eps^{(r)}, E_F^{(s)} \ll \eps^{(s)}, E_F^{(r)}$ so that for nearly each value of $k_y$, there exist electron-hole scattering channels.

The transport properties of the system are obtained from the scattering matrix $S$ that is calculated in the framework of the Landauer-B\"uttiker formalism using a variant of the recursive Green's function technique~\cite{PhysRevLett.47.882,*0022-3719-14-3-007,*springerlink:10.1007/BF01328846} we recently applied to a similar setup~\cite{PhysRevB.81.195441}. This technique is an efficient way to obtain the relevant parts of the system's Green's function from the surface Green's functions of the isolated leads, which is known analytically~\cite{PhysRevB.59.11936}, by solving Dyson's equation exactly, treating the coupling of the leads to the ring region as perturbation. The Fisher-Lee relation~\cite{PhysRevB.23.6851} then relates the Green's function to the $S$ matrix, from which the transmission function may be obtained. In this framework for elastic transport, Green's function and scattering matrix are parameterized by the eigenvalues $\eps$ of the Hamiltonian \eqref{for:hamBdG}.

In the following, we will concentrate on the regime $\eps < \Delta$, in which there are no propagating modes in the superconducting lead, so that electrons injected from the normal conducting lead are reflected back either as electron  ($e$) or hole ($h$). The scattering matrix thus has the structure
\begin{equation}
	S = \left(
	\begin{array}{cc}
		r_{ee}	& r_{eh} \\
		r_{he}	& r_{hh}
	\end{array}
	\right)
\end{equation}
from which the differential conductance for the Andreev processes is given by
\begin{equation}
	\frac{dI}{dV} = \frac{4e^2}{h} \, \mathrm{Tr}(r_{he}^\dagger r_{he})
\end{equation}
where the factor 4 accounts for spin degeneracy and the quantization of charge in units of $2e$.
\begin{figure}[t]
\centering
\includegraphics[width=0.95\columnwidth]{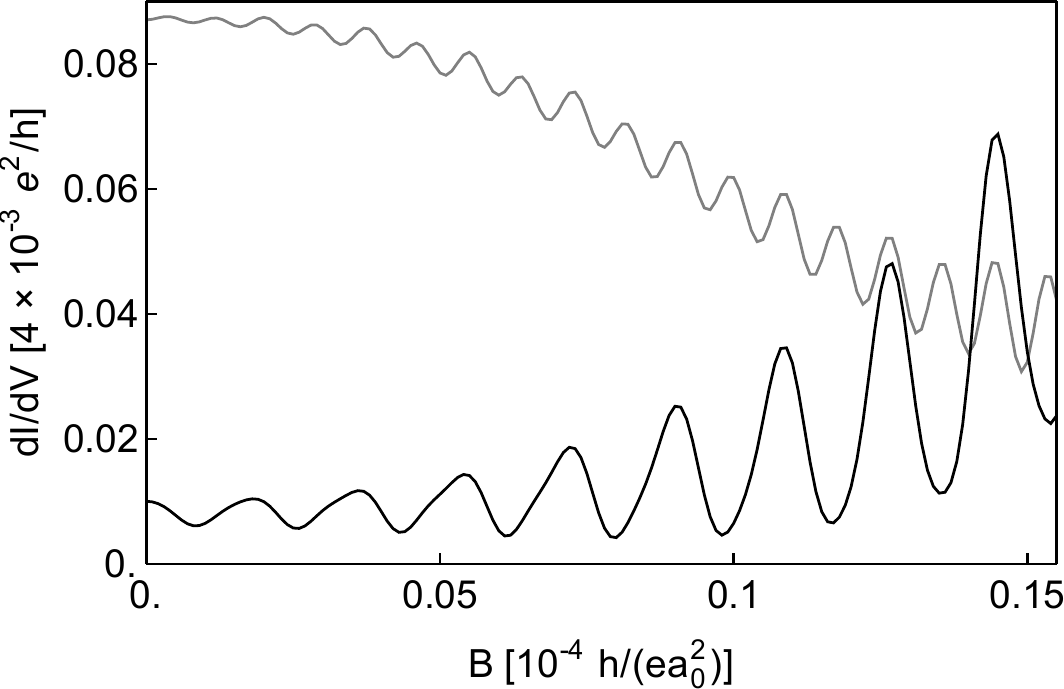}
\caption{Differential magnetoconductance for specular (black) and retro (gray) configuration for $E_F^{(r)} = 0.025\tau_0 = \eps^{(s)}$, $E_F^{(s)} = 0.001\tau_0 = \eps^{(r)}$, corresponding to 8 modes in the normal conducting lead, including all degeneracies (spin, valley, electron or hole). The high doping in the superconducting lead is chosen such that $E_F - U = 0.5\tau_0$ in both cases. Other parameter values are provided in the main text. The period of the dominant oscillation is $B_0^{(s)} \approx 1.8 \times 10^{-6} a_0^{-2} h/e$ in specular configuration and $B_0^{(r)} \approx 8.8 \times 10^{-7} a_0^{-2} h/e \approx 0.5 B_0^{(s)}$ in retroconfiguration. The weak beating pattern in retroconfiguration and the asymmetry in specular configuration arise due to minor contributions of contrary frequencies.}
\label{fig:results}
\end{figure}

In Fig.~\ref{fig:results}, we show the calculated transmission for a ring of width $w = 87 \sqrt{3} a_0$ and outer radius $R = 500 a_0$, where $a_0$ is the distance between nearest neighbors. The transmission function exhibits Aharonov-Bohm oscillations on top of a low frequency background which is due to universal conductance fluctuations. The position of the NS interface is given by $d = 400 a_0$. The chosen dimensions of the ring are large enough to exclude finite-size effects while still being numerically manageable. For the superconducting order parameter, we choose a value of $\Delta = 0.03 \tau_0 \approx 80$~meV, which may appear unrealistic at first sight, considered the fact that typical values are up to a few meV. However, by making this choice we scale the value of the superconducting order parameter according to the scale of the system size, such that the dimensionless factor $\Delta R / \hbar v_F$ stays of same order of magnitude, compared with values realized in experiments~\cite{PhysRevB.77.085413,1367-2630-12-4-043054}. 
Thus, for a realistic system size of $R \sim 10^{-6}$m, our choice of $\Delta$ would correspond to a value of a few meV for the superconducting gap. Note that due to these low energy scales and the rather large spacing of modes resulting from the narrow geometry of the electron waveguides in such a ring structure, in specular configuration only the regime of a low number of modes is accessible for realistic choices of system parameters. Also note that due to strong electron backscattering at the front of the hole and at the rough edges of the ring, the average value of the differential conductance is much less than a conductance quantum, $e^2/h$.

The average radius $\bar{r}$ of the scattering path is calculated according to $\bar{r}^2 \pi B_0 = h / n e$, where $n = 1$ ($n = 2$) in specular (retro) configuration and $B_0$ is the (dominant) period of the oscillation. Evaluating the period of the oscillations shown in Fig.~\ref{fig:results}, we obtain $\bar{r}^{(s)} \approx 420a_0$ in specular configuration and $\bar{r}^{(r)} \approx 425a_0$ in retroconfiguration. The obtained values lie well within the inner and outer radius of the ring and close to the arithmetic mean $R - w / 2 \approx 425a_0$, therefore confirming the predictions obtained from Eqs.~\eqref{for:Ts} and~\eqref{for:Tr}. Minor contributions of period $h/e$ in retroconfiguration and $h/2e$ in specular configuration visible in Fig.~\ref{fig:results} may arise due to terms neglected in Eqs.~\eqref{for:Ts} and ~\eqref{for:Tr}, scattering off the sharp boundaries of the ring structure, and the fact that for the electron-hole conversion at the NS interface $k_y$ is not strictly conserved.
\begin{figure}[t]
\centering
\includegraphics[width=0.95\columnwidth]{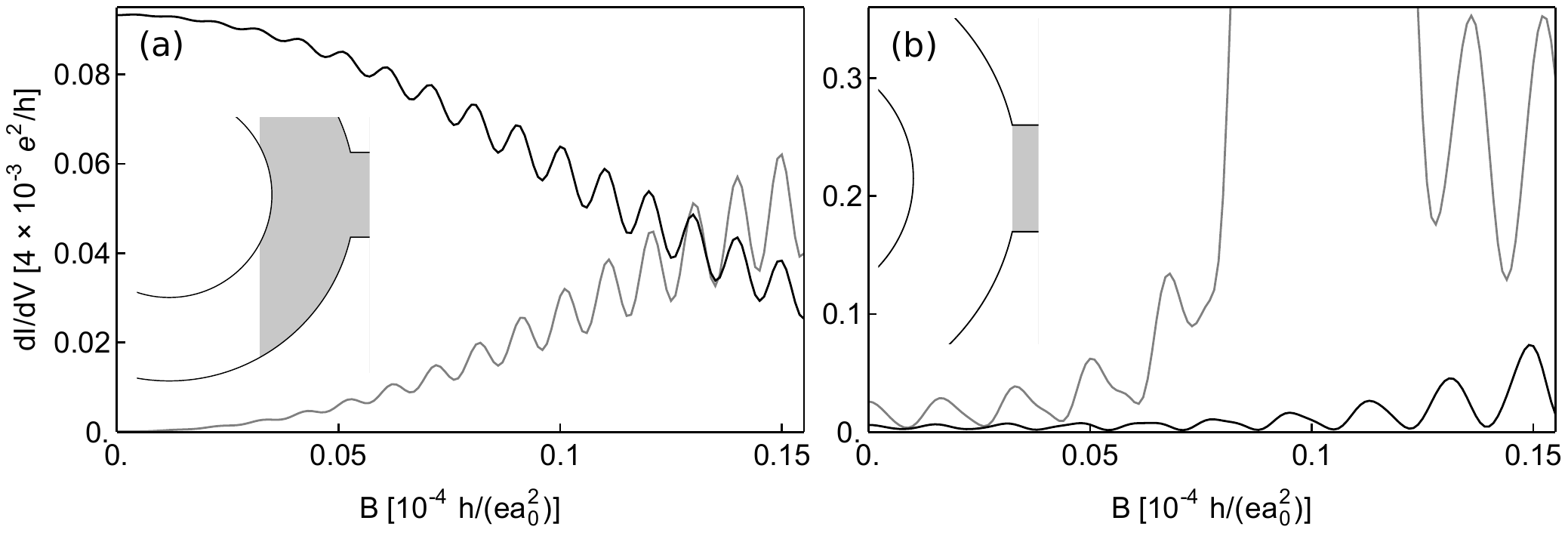}
\caption{Breakdown of the $h/e$ vs. $h/2e$ signature for shifted positions of the NS interface, as explained in the text. Other parameters and color coding are chosen as in Fig.~\ref{fig:results}. For $d = 340a_0$ (left), in specular configuration one observes oscillations of period $h/2e$ as in retroconfiguration. For $d = 490a_0$ (right), contributions of specularly reflected holes in retroconfiguration become important, leading to the observation of additional $h/e$-oscillations. The value of the superconducting coherence length is $\xi = 50a_0$.}
\label{fig:breakdown}
\end{figure}

Other strong evidence that supports our interpretation of the two different periods is the breakdown of this particular signature that is observed for a shift of the position of the NS interface on the scale of the width of the ring. Indeed, while in Ref.~\cite{PhysRevB.81.174523}---where a three-terminal graphene junction is analyzed---the exact position of the NS interface has no effect, it matters in our case; the reason is that $\xi$ is comparable or even less than the system size, while in Ref.~\cite{PhysRevB.81.174523} the superconducting coherence length greatly exceeds the system dimensions. If the interface is too close to the hole region [see Fig.~\ref{fig:breakdown}(a) inset], then specularly reflected holes are forced to traverse the same arm as the incoming electron. In this case, one should observe $h/2e$ oscillations in specular configuration. In contrast, if the interface is too far from the hole [see Fig.~\ref{fig:breakdown}(b) inset], holes may significantly be reflected through the other arm, e.\,g.\ due to increased scattering at the ring boundaries. This would manifest itself in the observation of $h/e$ oscillations in addition to the $h/2e$ oscillations in retroconfiguration. This behavior is confirmed in the observed magnetooscillations, as shown in Fig.~\ref{fig:breakdown}. 

\begin{figure}[b]
	\centering
	\includegraphics[width=0.95\columnwidth]{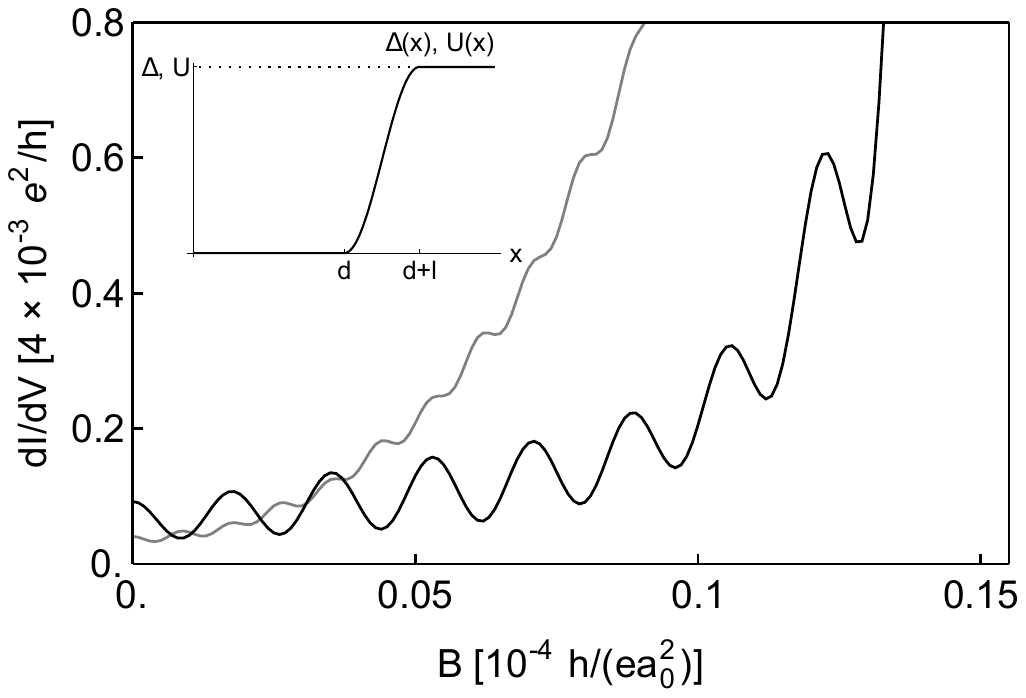}
	\caption{Magnetoconductance of the system used in Fig.~\ref{fig:results} with a smooth potential profile (inset) with $l = 90a_0$ and bulk disorder of strength $\sigma = 0.01\tau_0$ as explained in the text. The $h/e$ vs. $h/2e$ signature still persists. % 
	The color coding is the same as in Fig.~\ref{fig:results}.
}
	\label{fig:disorder}
\end{figure}

Apart from that, the $h/e$ vs. $h/2e$ signature proves to be very robust against moderate changes to the length and energy scales in the system, such as the extent of the magnetic field or the ratio of Fermi wavelength and the width of the NS interface. We also tested that the signature persists when more propagating modes are present in the lead, leading to values of the average conductance which are much larger as compared to the few-mode situation shown in Fig.~\ref{fig:results}. Additionally, the signature is hardly affected by bulk disorder, which is a major advantage of our setup. In Fig.~\ref{fig:disorder}, we show the magnetoconductance of the system used in Fig.~\ref{fig:results} with a particular random short-range disorder configuration, which 
is realized by applying an uncorrelated, random on-site potential of Gaussian distribution with zero mean and width $\sigma = 0.01\tau_0$ to each site. In addition, the NS interface has been smeared out over a distance $l = 90a_0$ in this case. The robustness of the effect can be explained from the topological nature of the signature: since all microscopic scattering paths  can be classified into just two groups---yielding $h/e$- or $h/2e$-oscillations, respectively---according to which arm is traversed by the quasiparticles, impurity scattering and the resulting deflection of quasiparticles has no adverse effect as long as scattering between the groups is weak, while scattering within one group may be arbitrarily strong. In addition, note that while our description of transport via the scattering matrix assumes complete phase coherence, a signature that distinguishes retro from specular Andreev reflection is assumed to persist also in the case of a finite phase coherence length. More specifically, if the phase coherence length is on the order of the ring circumference, first-order amplitudes in Eqs.~\eqref{for:Ts} and \eqref{for:Tr} may be neglected. Then, retroreflection would still manifest itself in $h/2e$-oscillations, while there would be no oscillations at all in the case of specular reflection.

Before we conclude, we like to add a remark concerning the choice of armchair boundary conditions in the leads we employed in our analysis. In a tight binding implementation of graphene, there are two simple choices for the orientation of the leads. Often, zigzag edges are considered to represent a generic boundary condition for graphene ribbons~\cite{PhysRevB.77.085423}. In this case, edge states are present in the system that modify the simple picture provided in Fig.~\ref{fig:dispersion} by adding additional scattering channels between bulk and edge states while removing certain scattering channels between bulk states due to the conservation of the so-called pseudoparity symmetry that acts like a selection rule~\cite{PhysRevB.79.115131}. In the realistic limit of metal leads providing a large number of propagating bulk modes, this effect %is not present. 
should be less important. 
However, for the system geometry used in the numerical calculations in combination with the low energy scales, it may significantly affect the observed behavior. In order to avoid this influence, we chose armchair boundary conditions in the leads that do not provide any edge states. Note in addition, that in realistic systems the zigzag-specific effect would also be suppressed since the zigzag edge state is not protected against disorder when next-nearest neighbor hopping is taken into account~\cite{epub12142}. Therefore, we are convinced that our results based on armchair edges in the reservoirs describe the generic situation for wide leads.

In conclusion, we have shown numerically that the frequency of Aharonov-Bohm oscillations in graphene rings provides a clear and feasible signature for distinguishing specular Andreev reflection from retroreflection. This feature can be explained qualitatively by considering the interference of the different scattering processes up to first order. The signature is robust against the presence of disorder and persists within a certain range for the position of the NS interface before it breaks down when the interface gets too close to or too far away from the hole of the ring.

We acknowledge interesting discussions with C.\,W.\,J.\ Beenakker, F.\ Dolcini, A.\,F.\ Morpurgo, and financial support from the European Science Foundation (ESF) under the EUROCORES Programme EuroGRAPHENE, the EU-FP7-project SE2ND, and the Emmy-Noether program of the DFG (P.\,R.).

%\bibliography{Bib}

%% Bibliography (contents of .bbl file as created by BibTeX)

\end{document}